\documentclass{phb-proc4-auth}

\usepackage{graphicx}
\usepackage{amssymb}

\begin{document}
\begin{frontmatter}


\journal{SCES '04}


\title{Nonlinear Transport in One-Dimensional Mott Insulator \\
in Strong Electric Fields}

%
%
%
%
%
%

\author{Takashi Oka\corauthref{1}}\author{, Ryotaro Arita and Hideo Aoki}

%
 
\address{Department of Physics, University of Tokyo, Hongo, Tokyo 113-0033, Japan}

%
%
%
%


%
%
%
%
\corauth[1]{Email:oka@cms.phys.s.u-tokyo.ac.jp}


\begin{abstract}
Time-dependent Schr\"{o}dinger's equation is integrated 
for a one-dimensional strongly-correlated 
electron system driven by large electric fields.  
For larger electric fields, many-body Landau-Zener tunneling 
takes place at anti-crossings of the many-body energy levels. 
The nonlinear $I$-$V$ characteristics as well as the 
time dependence of the energy expectation value are obtained. 
The energy of the Mott insulator in electric fields shows 
a saturation, which suggests a dynamical localization in energy space 
of many-body wave functions. 
\end{abstract}

%
%



\end{frontmatter}

%
%
%
%
%

{\it Introduction}
Strongly-correlated electron systems should be an 
interesting test-bench for nonequilibrium phenomena.  
Specifically, we can ask how the nonlinear transport properties 
of Mott insulators are distinct from those of band insulators. 
It has been experimentally shown that a quasi-one-dimensional 
cuprate exhibits a dielectric breakdown that remains when extrapolated 
to zero temperature\cite{tag}. This suggests that the origin of the
breakdown is quantum mechanical.  
Indeed, a many-body version of the Landau-Zener transition for 
correlated electron systems has 
been proposed by the present authors\cite{Oka2003}.  
In the present paper we make a further analysis of
the many-body Landau-Zener breakdown of the one-dimensional 
Mott insulator in strong electric fields to elucidate 
how its nature differs from those of the conventional 
Zener breakdown of band insulators.

{\it $I$-$V$ characteristics }
We study the time evolution of 
wave functions for a one-dimensional Mott insulator 
in a static electric field $F$ turned on at $t=0$.
The electric field is here induced by 
a time-dependent Aharonov-Bohm(AB) flux $\phi(t)=\Phi(t)/\Phi_0=FLt/h$ 
pierced through the system in a periodic boundary condition, 
so the evolution is governed by the 
time-dependent Schr\"odinger equation 
with a time-dependent Hamiltonian,
\begin{eqnarray}
H(\phi(t))&=&-\frac{w}{4}
\sum_{i\sigma}\left(e^{i2\pi\phi(t)/L}c_{i+1\sigma}^\dagger 
c_{i\sigma}+{\rm H.c.}\right)\\
&&+ U\sum_in_{i\uparrow}n_{i\downarrow} 
+\frac{\Delta}{2}\sum_i(-1)^in_i.\nonumber
\end{eqnarray}
When the system is half-filled, 
the groundstate is a band insulator for the 
level offset $\Delta>0$ with the Hubbard $U=0$, 
and becomes a Mott insulating when $U>0$ with $\Delta =0$.

\begin{figure}[ht]
\centering 
\includegraphics[width= 6.cm]{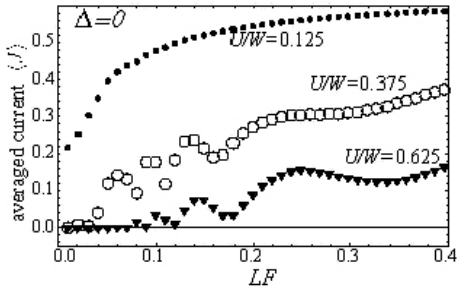}
\caption{
$I$-$V$ characteristics for the one-dimensional 
Hubbard model for an $N=6$ site ring.
}
\label{IV}
\end{figure}
We first look at the
$I$-$V$ characteristics (Fig.\ref{IV}) of the
Mott insulating phase ($U>0,\;\Delta =0$), 
where the current is defined as a time-averaged 
$\langle J\rangle \equiv\frac{1}{T}\int_0^T\langle J(t)\rangle dt$.
The current is suppressed until the field strength exceeds
a threshold, $\overline{F}(U)$, whose value depends on $U$. 
The threshold can be estimated from the Landau-Zener formula,
and we predict an $\overline{F}(U)\propto [\Delta E (U)]^2$ 
dependence on the charge gap $\Delta E (U)$ 
as confirmed numerically\cite{Oka2003}.
Above the threshold, Landau-Zener tunneling from the 
ground state to excited states is activated.

\begin{figure}[ht]
\centering 
\includegraphics[width= 8.cm]{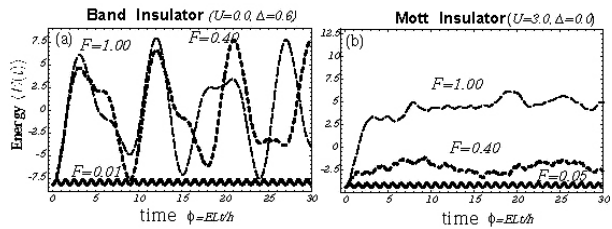}
\caption{Time evolution of the energy $\langle H(t)\rangle$
of a band insulator(a) and a Mott insulator(b).  
}
\label{Energy}
\end{figure}

What distinguishes the breakdown of Mott insulators 
from those for band insulators is how 
the system becomes out of equilibrium 
in electric fields above the threshold. 
In order to quantify this, we plot the long-time behavior of the
energy expectation value $\langle H(t)\rangle$ 
for a Mott insulator as compared with that for a
band inslator(Fig.\ref{Energy}). 
\begin{description}
	\item[Small $F$ regime:] 
    When the Landau-Zener tunneling is ineffective, the wave function 
    approximately evolves adiabatically along the 
    ground state of the time-dependent Hamiltonian $H(\phi(t))$. 
    The ground-state energy shows an AB-oscillation for a finite system.
    \item[Large $F$ regime:]
    $\langle H(t)\rangle$ of band insulators diverges 
    in the long-time limit, 
    which is not apparent in Fig.\ref{Energy}(a), 
    but the long-period oscillation 
    seen in the figure has a periodicity $\Delta \phi \sim N$, 
    so the amplitude as well as the periodicity 
    diverge for $N\rightarrow \infty$.
    This is in sharp contrast with the 
    saturated behavior of $\langle H(t)\rangle$ of Mott insulators.
\end{description}

The divergence of $\langle H(t)\rangle$ in the 
band inslator can be readily understood. The Landau-Zener tunneling produces 
free electron-hole pairs, which are accelerated 
by static electric fields.  Since there is no dissipation 
for the present clean and isolated system 
nor electrode to absorb the energy, the energy increases 
indefinitely. 

\begin{figure}[ht]
\centering 
\includegraphics[width= 8.cm]{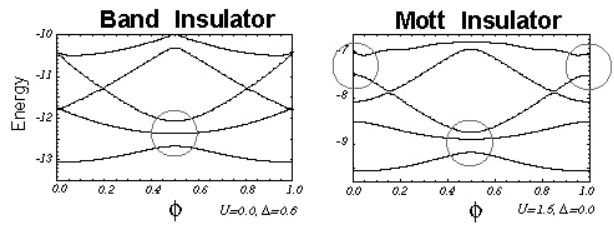}
\caption{The adiabatic level scheme for a 
band insulator(a) and a Mott insulator (b). 
The level anticrossings are highlighted. }
\label{spec}
\end{figure}

On the other hand, $\langle H(t)\rangle$ 
in the Mott insulator is seen to saturates to a value 
that depends on the electric field, which is quite an 
interesting and nontrivial behavior.  
For noninteracting but disordered systems, 
Gefen and Thouless explained such a saturation in terms of a 
{\it dynamical localization}, i.e., Anderson's localization in 
energy space\cite{Gefen1987}.  We can extend the notion of the dynamical 
localization to a (clean) many-body system.\cite{Konno2004}  
This takes place on many-body adiabatic spectra 
typically depicted in Fig.\ref{spec}.  
We can see that, while a band insulator has a single level anticrossing 
separated by a gap between the ground state 
and the state excited with one electron-hole pair, 
a Mott insulator has many (actually 
a macroscopic number of) anticrossings among the excited 
levels.  At each level anticrossing the Landau-Zener tunneling takes 
place and the probability amplitude bifurcates. 

An important point to note is that this branching occurs in a 
quantum mechanical manner, where contributions, with their phase factors, from 
various paths (in the energy-level vs time space) interfere 
with each other, resulting in the dynamical localization.  
The amplitude of the distribution resides near the 
ground state when the electric field is weak, 
whereas a component extending into excitated states arises 
when $F$ becomes larger, which is governed by the above phase factor.

{\it Conclusion}

So, while the disorder gives rise to a saturated 
behavior in the current-time characteristics in one-body 
systems, this is replaced by the quantum mechanical tunneling across 
many level repulsions caused by the electron-electron 
interaction in many-body systems.  
We have identified this as the reason why the 
strong field behaviors are 
quite different between band insulators and
Mott insulators.  

Part of this work was supported by a Grant-in-Aid for JSPS
fellows for Young Scientists, from the 
Japanese Ministry of Education.

%
%
%
%

%
%
%
%


\end{document}